\documentclass[aps,pre,amsmath,floatfix,superscriptaddress,color]{revtex4}
\usepackage{bm}% bold math
\usepackage{amsmath} % \pmb bold special characters
\usepackage[bitstream-charter]{mathdesign} % TB
\usepackage{rotating}

\RequirePackage{color}

\def\mat#1{{\mathbf#1}}
\def\nbOne{\pmb{\mathsf 1}}
\def\vecc#1{\vec{#1}}

\def\modrm#1{{#1}}

\def\naA{{\mathsf W}}
\def\naAmat{\pmb{\mathsf W}}
\def\naGmat{\pmb{\mathsf G}}
\def\naFmat{\pmb{\mathsf F}}
\def\naE{{\mathsf E}}
\def\naEmat{\pmb{\mathsf E}}
\def\naB{{\mathsf B}}
\def\naBmat{\pmb{\mathsf B}}
\def\naphi{{\chi}}
\def\naphimat{\pmb{\chi}}
\def\nadermat{{\pmb{\cal D}}}
\def\bsigma{{\pmb{\sigma}}}

\def\rhospin{s}
\font\fontedimension=eufm10
\def\di#1{\hbox{\fontedimension #1}}

\begin{document}
%\jl{31}
\title{\sffamily \bfseries 
	Classical Yang-Mills theory in condensed matter physics}

\author{Bertrand Berche}
\affiliation{Statistical Physics Group, P2M, Institut Jean Lamour, 
	UMR CNRS 7198, 
	Nancy Universit\'e, BP70239, F-54506 Vand\oe uvre les Nancy, France}
\affiliation{Laboratorio de F\'\i sica Estad\'\i stica de Medios Desordenados,
 Centro de F\'\i sica, Instituto Venezolano de Investigaciones 
	Cient\'ificas, Apartado 21874, Caracas 1020-A, Venezuela}

\author{Ernesto Medina}
\affiliation{Laboratorio de F\'\i sica Estad\'\i stica de Medios Desordenados,
 Centro de F\'\i sica, Instituto Venezolano de Investigaciones 
	Cient\'ificas, Apartado 21874, Caracas 1020-A, Venezuela}
\affiliation{Statistical Physics Group, P2M, Institut Jean Lamour, 
	UMR CNRS 7198, 
	Nancy Universit\'e, BP70239, F-54506 Vand\oe uvre les Nancy, France}

\vskip10mm
%\date{\today}

\begin{abstract}
\baselineskip=9pt
Recently, gauge field theory approaches were extensively used
in order to discuss the physical 
consequences of spin-orbit interactions in condensed matter physics. 
An SU(2)$\times$U(1) gauge theory is very naturally borne out and provides
an illustrative example of a classical Yang-Mills field theory at work. 
This approach may serve as an exemplification of non-Abelian field theories 
for students in general physics curriculum. It allows to introduce
discussions on fundamental ideas like Noether currents, gauge symmetry 
principle, gauge symmetry breaking and non linear 
Yang Mills equations in very concrete physical situations {that makes it accessible
to a broad audience.}
\\
\\
{\bf R\'esum\'e.} 
Les th\'eories de jauge ont r\'ecemment \'et\'e utilis\'ees de mani\`ere syst\'ematique pour discuter les cons\'equences
physiques de l'interaction spin-orbite en physique de la mati\`ere condens\'ee. Une th\'eorie SU(2)$\times$ U(1) appara\^it naturellement
et fournit un exemple d'illustration du fonctionnement des th\'eories classiques de Yang et Mills. Cette approche peut servir d'example de th\'eorie
de jauge non ab\'elienne pour des \'etudiants dans un cours de physique g\'en\'erale. Cela permet d'introduire et de discuter les id\'ees fondamentales
comme les courants conserv\'es de Noether, le principe de sym\'etrie de jauge, la brisure de la sym\'etrie de jauge et les \'equations non lin\'eaires de Yang et Mills dans des situations physiques concr\`etes, les rendant ainsi accessible \`a une vaste communaut\'e.

{\footnotesize
$$
\begin{array}{ll}
\varphi(\vec r,t) &  \hbox{wave function or complex scalar field}\\
\psi(\vec r,t) &  \hbox{Pauli $2-$component spinor or Pauli spinor field}\\
\Psi(\vec r,t) &  \hbox{Dirac $4-$component spinor}\\
{\cal L}  &  \hbox{Lagrangian density} \\
 H &  \hbox{Schr\"odinger Hamiltonian} \\
 \mat H  &  \hbox{Pauli Hamiltonian} \\
{\cal D}_t,\ \vec{\cal D} & \hbox{U(1) covariant derivatives}\\
\phi,\ \vec A & 
	\hbox{U(1) scalar and vector gauge fields}\\
\vec E,\ \vec B & 
	\hbox{electric and magnetic fields}\\
\rho(\vec r,t)  &  \hbox{charge density} \\
\vec j(\vec r,t)\ \hbox{or}\ \vec J(\vec r,t) &  
	\hbox{charge current density} \\
e,\ g & \hbox{electric and isospin charge}\\
{\boldsymbol \tau}^a & \hbox{SU(2) generators}\\
{\boldsymbol\alpha} & 
	\hbox{SU(2) contraction ${\boldsymbol\tau}^a\alpha^a$}\\
{\nadermat}_t,\ \vec{\nadermat} & 
	\hbox{SU(2) covariant derivatives}\\
{\bf s}^a & \hbox{$a-$component of the spin $\frac 12$ operator}\\
{\boldsymbol \sigma}^a & \hbox{$a-$ component Pauli matrix}\\
\naphimat=\frac12{\boldsymbol\sigma}^a\naphi^a,\ 
\vec\naAmat=\frac12{\boldsymbol\sigma}^a\vec\naA^a & 
	\hbox{SU(2) scalar and vector gauge fields}\\
\vec\naEmat=\frac12{\boldsymbol\sigma}^a\vec\naE^a,\ 
\vec\naBmat=\frac12{\boldsymbol\sigma}^a\vec\naB^a & 
	\hbox{SU(2)  field strengths}\\
s^a_{\rm mat.}(\vec r,t),\ \rho^a_{\rm mat.}(\vec r,t),\ \vec J^a(\vec r,t)  &  
	\hbox{$a-$component of the matter spin or isospin density and current 
	density} \\
s^a_{\rm rad.}(\vec r,t),\ \rho^a_{\rm rad.}(\vec r,t),\ \vec {\cal J}^a(\vec r,t)  &  
	\hbox{$a-$component of the radiation 
	spin or isospin density and current 
	density} \\
\vec s(\vec r,t)  &  
	\hbox{vector spin density} \\
\end{array}
$$
}

\end{abstract}

\pacs{\\ 
75.76.+j: 	Spin transport effects\\
	11.15.-q: 	Gauge field theories\\
	75.70.Tj: 	Spin-orbit effects}

\ \hskip2.37cm\today

%\maketitle
%%%%%%%%% INTRODUCTION %%%%%%%%%
%\documentclass[aps,pre,amsmath,floatfix,superscriptaddress,color]{revtex4}

\maketitle

\section{Introduction}

Yang-Mills theories are usually introduced 
to students in the context of quantum field theory, but are never encountered 
in their classical form at the undergraduate level~\cite{Boozer11}.
{Regarding the fundamental role of gauge symmetry in the construction of
physical theories, this is desirable in order to make the underlying concepts
accessible to physics students early in the curriculum.} 
In the abstract of a paper by R. 
Mills~\cite{Mills89}, the content of gauge theories 
is summarized as follows: {\em  
``\dots the gradual emergence of symmetry as a driving force
	in the shaping of physical theory; the elevation of Noether's theorem,
	relating symmetries to conservation laws, to a fundamental principle 
	of nature; and the force idea (``the gauge principle'') that the 
	symmetries of nature, like the interactions themselves, should be 
	local in character.''}
{This obviously brings the concept of gauge invariance to the level of a
keystone principle among physical theories, rather than a specific property of 
electromagnetic theory.}

The non relativistic limit of Dirac equation, namely the Pauli equation and its
corrections to Schr\"odinger theory, i.e. the spin-orbit (SO)
interaction and the Zeeman interaction, offers 
a natural frame for an SU(2) gauge field theory. This does not only give 
an elegant formulation for the coupling of spin degrees of freedom with external
electromagnetic fields, but also the Yang-Mills approach provides natural
answers to fundamental issues such as conservation laws which may offer
 illuminating insights in the condensed matter physics context~\cite{Sonin10}. 
Here we show that it may be a useful
and very concrete way to introduce Yang-Mills theory, in the condensed matter physics curriculum.

A reformulation of the spin-orbit coupling Hamiltonian in terms of 
non-Abelian gauge fields~\cite{YangMills54,Utiyama,Dotson66} 
was explicitly given in 
ref.~\cite{Rebei,Jin,Leurs,Medina,BercheEtAl09,DartoraCabrera10,FujitaEtAl2011} where the 
SO interaction 
is presented 
as a  SU(2)$\times$ U(1) gauge theory. A key property for that purpose is to 
deal with an SO interaction which is linear in the momentum, 
$H_{SO}\sim{\vecc p}\cdot(\vec{\boldsymbol \sigma}\times{\vecc E})$ in such 
a way that it can be absorbed in the kinetic energy through a minimal 
coupling procedure, ${\vecc p}\longrightarrow{\vecc p}
-\vec{\boldsymbol \sigma}\times{\vecc E}$, up to unessential numerical 
factors. This property is fulfilled by the Pauli Hamiltonian, but also by
the Rashba
SO interaction~\cite{Rashba} and by the two-dimensional reduction of the 
Dresselhaus SO
interaction~\cite{Dresselhaus}. 
A potential realization of such situation in the general case
is thus the $2-$dimensional electron gas (2DEG) as encountered in 
semiconductor hetero-junctions~\cite{Winkler} and 2DEG quantum rings~\cite{BercheChatelainMedina}.

Such gauge point of view, in more general terms, has been known for some 
time\cite{Goldhaber,Mineev,Frohlich}. This formulation is very revealing, 
since the consistent gauge structure of the theory becomes obvious and the 
physics of spin currents, persistent currents and color 
diamagnetism\cite{Tokatly} can be understood in a manner analogous to the 
well known U(1) gauge theories. 

{The plan of the paper is as follows}: we start in section~\ref{secII} with
the case of electromagnetic theory which will then serve as a basis to the
non-Abelian {Yang-Mills theory presented} in section~\ref{secIII}.
Section~\ref{secIV} discusses, in this context, the case of spin-orbit and Zeeman
interactions and application to Einstein-de Haas effect
is discussed in section~\ref{secVI}.
We have decided to avoid most of the time the use of tensor analysis and instead embrace
vector algebra, in order to obtain equations of motion in a form very
similar to Maxwell equations in matter, as they are known in standard
textbooks~\cite{PanofskyPhillips}. 
We feel that this will help the part of the
community who is less familiar with field theory, appealing to the
broader intuition from electromagnetism.

\section{A well-known gauge field theory: the example of electromagnetism}
\label{secII}
In this section, we collect the standard results of classical
gauge field theory applied to electromagnetism in order to {later follow} a
parallel approach in the case of non Abelian and SO interaction. 

\subsection{Conserved currents}

A fundamental question in physics poses the problem of the electric charge.
{\em Why is charge
 conserved?}
Historically we have made the empirical observation that in all
fundamental processes, e.g. collisions between elementary particles, the
electric charge is conserved. 
The sum of the charges
of the outgoing particles is always equal to the
sum of the charges of the incoming particles. This property then translates
into a hydrodynamic conservation equation $\partial_t\rho(\vec r,t)
+\vec\nabla\cdot\vec j(\vec r,t)=0$, where $\rho(\vec r,t)$ is the
electric charge density calculated on a coarsed grained scale when 
a continuous density make sense. Integration over a mesoscopic 
(or macroscopic) scale, then leads to $\frac{dQ(t)}{dt}=
\int d^3r\ \!\partial_t\rho(\vec r,t)=-\oint\vec j(\vec r,t)\cdot d\vec S=0$
provided that the volume of integration is large enough and the current 
densities vanish at the boundaries. So conservation of charge follows from
an experimental observation, but its origin is not something obvious.
An answer to the question of the origin of such a conservation law (formulated here in a provocative
manner) is that Hilbert spaces and
the complex fields living in these spaces don't associate physical relevance to 
global phases of such fields. Here we refer to the fact that in 
Quantum Mechanics (QM) the results of measurements are given by the
{matrix elements  $\langle\varphi|Obs.|\varphi\rangle$
of Hermitian operators $Obs.$ associated to physical observables, and such
matrix elements are independent of the choice of reference for the phase
of the complex vector $\varphi$.} 

The principle of conservation of the electric charge, experimentally 
legitimate, is derived {using} the second approach from a more
fundamental and more abstract principle of invariance of the theory with 
{respect to global phase changes}.
The main advantage of the second answer is that it enables us to connect to other
similar theories. More importantly, it leads to a consistent construction of the
electromagnetic theory with {\em no a priori knowledge} other than that implied by
the Lagrangian formalism which is the language adapted for such approaches. This
is what we now show.

The Schr\"odinger equation follows from a least action principle~\cite{Doughty}
\begin{equation}\frac\delta{\delta\varphi^*}\int 
{\cal L}_0 d^3r\ \! dt
=\frac{\partial {\cal L}_0}{\partial\varphi^*}-\partial_k
\frac{\partial {\cal L}_0}{\partial(\partial_k\varphi^*)}
-\partial_t
\frac{\partial {\cal L}_0}{\partial(\partial_t{\varphi^*})}=
0\label{EL0}
\end{equation}
where the free matter Lagrangian density ${\cal L}_0$ 
is an ordinary function ${\cal F}$ of
the wave function $\varphi(\vec r,t)$
(a complex scalar field describing matter), its space and time derivatives, and the
corresponding complex conjugates treated as independent variables.
${\cal L}_0={\cal F}(\varphi,\varphi^*,\vec\nabla\varphi,\vec\nabla\varphi^*,
\partial_t\varphi,\partial_t\varphi^*)$ is 
the Jordan-Wigner Lagrangian density~\cite{Doughty,BrownHolland}, 
\begin{equation}{\cal L}_0=
i\hbar\varphi^*\partial_t\varphi-\frac{\hbar^2}{2m}(\vec{{\nabla}}\varphi)^*
(\vec{{\nabla}}\varphi)-V\varphi^*\varphi\label{JWLag}\end{equation}
and equation~(\ref{EL0}) applied to this Lagrangian
recovers the Schr\"odinger equation.
As discussed above, Quantum Mechanics is invariant 
under a {\em global} change
of phase of the wave function 
\begin{equation}\varphi({\vecc r},t)\to\exp(ie\alpha/\hbar)\varphi({\vecc r},t),\quad\alpha={\rm const.}\end{equation} 
since all physical
quantities, expressed in terms of matrix elements, are independent
of the global phase choice $\alpha$. The coefficient 
$e/\hbar$ in the phase is present for later convenience. It simply fixes
the dimensions of the $e\alpha$ to those of an action.
The phase transformation above is a continuous 
symmetry ($\alpha$ can take any real value) which leaves the Lagrangian
density unchanged.
Applying Noether theorem to an infinitesimal gauge transformation $\varphi\to\varphi'\simeq (1+ie\alpha/\hbar)\varphi=\varphi+\delta\varphi$~\cite{Doughty,Noether}, 
 it follows that we can define a 
conserved current
obeying a continuity equation. Indeed, demanding the global gauge transformation to be a symmetry, i.e. to preserve the lagrangian density
${\cal L}_0$ unchanged leads to 
% \begin{widetext}
\begin{eqnarray}
\hskip-15mm
\delta{\cal L}_0=0=
\frac{\partial{\cal L}_0}{\partial\varphi}\delta\varphi+
\frac{\partial{\cal L}_0}{\partial\dot\varphi}\delta\dot\varphi+
\frac{\partial{\cal L}_0}{\partial\vec\nabla\varphi}\delta\vec\nabla\varphi
+(\varphi\to\varphi^*)
&&
\nonumber\\
&&\hskip-80mm
=-\frac{ie\alpha}\hbar\left[\partial_t\left(\varphi^*\frac{\partial{\cal L}_0}{\partial\dot\varphi^*}-\frac{\partial{\cal L}_0}{\partial\dot\varphi}\varphi\right)+\vec\nabla\cdot\left(\varphi^*\frac{\partial{\cal L}_0}{\partial\vec\nabla\varphi^*}-\frac{\partial{\cal L}_0}{\partial\vec\nabla\varphi}\varphi\right)\right]\nonumber\\
&&\hskip-80mm
=-\alpha(\partial_t\rho+\vec\nabla\cdot{\vecc j})\label{eq-continuity-j}\end{eqnarray}% \end{widetext}
where use has been made of the equations of motion (\ref{EL0}) in the second line.
This leads to the definition of a {\em charge density} $\rho({\vecc r},t)=
e\varphi^*({\vecc r},t)\varphi({\vecc r},t)$ and a {\em charge current 
density} ${\vecc j}({\vecc r},t)=
\frac{-ie\hbar}{2m}(\varphi^*({\vecc r},t)\vec\nabla\varphi({\vecc r},t)
-\varphi({\vecc r},t)\vec\nabla\varphi^*({\vecc r},t))$. This very
interesting connection
between ``phase invariance'' in QM and charge conservation was first made
by F. London. For interested readers, historical aspects of gauge invariance
are  discussed at length in the book of O'Raifeartaigh~\cite{ORaifeartaigh}
or in the reviews~\cite{ORaifeartaigh2,WuYang06}. 

\subsection{The gauge invariance principle illustrated}

Let us now quote Salam and Ward~\cite{SalamWard1961}, 
(see also Novaes~\cite{Novaes}), who speculated {on whether} a local
generalization of the gauge invariance would generate strong, weak and electromagnetic interactions and {the} associated new fundamental conservation laws:
{\em ``Our basic postulate is that it should be possible to generate
strong, weak, and electromagnetic interaction terms ... by making
local gauge transformations on the kinetic-energy terms in the free
Lagrangian for all particles.''} The message behind this sentence is that
theories which have a larger symmetry class (local gauge symmetry)
have richer dynamical structure (new interactions present), and 
the power of gauge field theory appears when one assumes that the phase 
invariance invoked above also holds in the case of {\em local} phase
changes $\alpha({\vecc r},t)$,
\begin{equation}\varphi({\vecc r},t)\to\exp(ie\alpha({\vecc r},t)/\hbar)\varphi({\vecc r},t)\equiv G\varphi({\vecc r},t),\label{eq-localgaugeinv}\end{equation} 
where $G\equiv G({\vecc r},t)$.
This is obviously not a symmetry of the original Lagrangian 
density ${\cal L}_0$, since $\vec\nabla\varphi({\vecc r},t)$ and $\partial_t
\varphi({\vecc r},t)$ 
do not transform like the wave function itself, and as a consequence, terms like
$i\hbar\varphi^*\partial_t\varphi$ or $-\frac{\hbar^2}{2m}\vec\nabla\varphi^*
\vec\nabla\varphi$ are not gauge invariant (the first of these terms for
example transforms into
$i\hbar\varphi^*\partial_t\varphi
-e(\partial_t\alpha)\varphi^*\varphi$). 
In order to repair this lack of symmetry of the original Lagrangian, and to be
able to build a new one with the expected invariance property, one may define
so-called {\em covariant} derivatives 
$\vec{\cal D}=\vec\nabla-\frac{ie}{\hbar}{\vecc A}({\vecc r},t)$ 
and ${\cal D}_t
=\partial_t+\frac{ie}{\hbar}\phi({\vecc r},t)$ in such a way that acting on 
$\varphi(\vec r,t)$, they
transform like the wave function, i.e. according to
$\vec{\cal D}\varphi({\vecc r},t)
\to\exp(ie\alpha({\vecc r},t)/\hbar)\vec{\cal D}\varphi({\vecc r},t)$ and 
${\cal D}_t\varphi({\vecc r},t)
\to\exp(ie\alpha({\vecc r},t)/\hbar){\cal D}_t\varphi({\vecc r},t)$. 
The two fields
${\vecc A}({\vecc r},t)$ and $\phi({\vecc r},t)$ are introduced
for the transformations above to be satisfied. Again, the factors
$e/\hbar$ are present for convenience and the imaginary $i$ is a choice
which enables $ \phi$ and $\vec A$ to be real fields (we require 
$i\partial_t$ and $-i\vec\nabla$ to be Hermitian operators).  
The gauge
transformation conditions
imply that these fields obey the laws
\begin{eqnarray}
{\vecc A}({\vecc r},t)&\to& G\vec A({\vecc r},t)G^{-1}
+\frac \hbar{ie}(\vec\nabla G)G^{-1}%\nonumber\\&&
=
{\vecc A}({\vecc r},t)
+\vec\nabla\alpha({\vecc r},t)\\
\phi({\vecc r},t)&\to& G\phi({\vecc r},t)G^{-1}
-\frac \hbar{ie}(\partial_t G)G^{-1}%\nonumber\\&&
=
\phi({\vecc r},t)-\partial_t\alpha({\vecc r},t),\end{eqnarray} where one 
recognizes the gauge tranformations of the vector and scalar potentials
of electromagnetism.%~\cite{Kobe}. 

Summarizing the discussion {to this point}, we made a local phase change to the wave function, but this does not
leave the Lagrangian density invariant unless we compensate this
change i.e. the terms brought about by derivating $\alpha$ can be
cancelled by corresponding changes in the gauge fields. The terminology of 
covariant derivative refers to the fact that ${\cal D}_t\varphi$ and $\vec{\cal D}\varphi$ are not invariant under a gauge transformation, but do transform
in the same manner as the field $\varphi$ itself. The derivative operators ${\cal D}_t$ and $\vec{\cal D}$ themselve transform like vectors under a
gauge transformation, ${\cal D}_t\to G{\cal D}_tG^{-1}$ and $\vec{\cal D}\to G\vec{\cal D}G^{-1}$.
In the presence of an external gauge field, one way to make the Lagrangian density 
gauge invariant is the {\em minimal coupling 
prescription}, i.e. replace the ordinary derivatives by the covariant ones in the
free matter field Lagrangian density~(\ref{JWLag}),
\begin{eqnarray}
{\cal L}_{\rm mat.} &\equiv&
{\cal F}(\varphi,\varphi^*,\vec{\cal D}\varphi,
(\vec{\cal D}\varphi)^*,
{\cal D}_t\varphi,({\cal D}_t\varphi)^*)\nonumber\\
&=&{\cal L}_{0}(\varphi,\varphi^*,\dots)+{\cal L}_{\rm int.}(\varphi,\varphi^*,\dots ,\phi,{\vecc A}),
\end{eqnarray}
where now there appears a supplementary  
interaction term, depending on both the ``matter field'' $\varphi$ and the
gauge fields $\phi$, $\vec A$. This term appears because any term allowed by the
gauge symmetry should appear. In principle higher order terms can also appear but one takes the minimal lowest orders. The {price to pay} to extend the
so-called U(1) 
gauge invariance to local phase transformations is that it automatically
generates the interactions of the charged matter field with the 
electromagnetic field, {which is a strong self consistent check for the theory!}
Yang summarized the gauge invariance approach to the study of fundamental 
interactions by the expression 
{\em  ``Symmetry dictates interaction''}~\cite{CNYang}
%\revision{this couples well with the principle above that one should include the cross term because it obeys symmetry}. 

In terms of the gauge fields, the {\em physical} electric
and magnetic fields $\vecc E$ and $\vecc B$ are defined
as the commutators of covariant derivatives, 
${\modrm E}_k=\frac{\hbar}{ie}[{\cal D}_t,{\cal D}_k]$ and
${\modrm B}_k=-\frac 12\frac{\hbar}{ie}\epsilon_{ijk}[{\cal D}_i,{\cal D}_j]$
($\epsilon_{ijk}$ is the totally antisymmetric tensor which takes
value $+1$ when $ijk$ is $123$ up to circular permutations, it is
$-1$ when $ijk$ is $213$ up to circular permutations, and it is $0$
when two or three indices are identical). In $4-$dimensional notation, one introduces the
Faraday tensor $F_{\mu\nu}=\frac\hbar{ie}[{\cal D}_\mu,{\cal D}_\nu]$ (with 
${\cal D}_\mu=\partial_\mu+\frac{ie}\hbar A_\mu$ and the usual metric signature, $A_\mu = (\phi/c,-\vec A)$) which 
transforms under gauge transformation as a vector, i.e. $F_{\mu\nu}\to GF_{\mu\nu}G^{-1}$. The Faraday tensor (hence the
electric and magnetic fields) is thus gauge invariant in the Abelian U(1) gauge theory. The contribution of the free gauge fields
to the Lagrangian density then follows simply from the following requirements: we demand a i) Lorentz invariant, ii) quadratic in the gauge fields
derivatives (i.e. a kinetic energy of a standard form) and iii) gauge invariant quantity. Prescription i) is fulfilled if all space-time indices
are contracted while ii) is nicely obtained from the square of the Faraday tensor. The quantity $F_{\mu\nu}F^{\mu\nu}$, which transforms
under gauge transformation as $F_{\mu\nu}F^{\mu\nu}\to GF_{\mu\nu}F^{\mu\nu}G^{-1}$ {satisfies} all three conditions, since $G$ and
$G^{-1}$ simplify each other in the present Abelian theory for which all quantities commute.
This free gauge field
Lagrangian density is known as  the Schwarzschild {Lagrangian} 
density~\cite{Schwarzschild,PalSateesh90,Essen09}),%,HuangLin2002}),
\begin{eqnarray}
{\cal L}_{U(1)}&=&-{\scriptstyle\frac 1{4\mu_0}}F_{\mu\nu}F^{\mu\nu}%\nonumber\\&=&
={\scriptstyle\frac 12}\varepsilon_0{\vecc E}^2-
{\scriptstyle\frac{1}{2\mu_0}}{\vecc B}^2
%\nonumber\\
%&=&{\scriptstyle\frac 12}\varepsilon_0(-\vec\nabla\phi-\partial_t{\vecc A})^2
%-{\scriptstyle\frac 1{2\mu_0}}(\vec\nabla\times{\vecc A})^2
.
\end{eqnarray}

The formulation of the principle of gauge invariance was stated in Yang and Mills
work~\cite{YangMills54}, where they generalize ``phase invariance'' of ordinary Quantum Mechanics
of charged fields. {They always refer} to the case of
electromagnetic theory in order to illustrate their {arguments}:
{\em We define {\it isotopic gauge} as an arbitrary way of choosing the orientation of 
the isotopic spin axes at all space-time points, in analogy with the electromagnetic gauge
which represents an arbitrary way of choosing the complex phase factor of a charged field
at all space-time points. We then propose that all physical processes (not involving the electromagnetic field) be invariant under the isotopic gauge transformation $\psi\to\psi'$, 
$\psi'=S^{-1}\psi$, where $S$ represents a space-time dependent isotopic spin rotation.

To preserve invariance one notices that in electrodynamics it is necessary to counteract the 
variation of $\alpha$ {\rm [the $\alpha$ of Yang and Mills is the same as our $-e\alpha$ in Eq.~(\ref{eq-localgaugeinv})]} with $x$, $y$, $z$ and $t$ by introducing the electromagnetic field $A_\mu$
which changes under a gauge transformation as 
$$A'_\mu\to A_\mu+\frac 1e\partial_\mu\alpha.$$
In an entirely similar manner one introduces a $B$ field in the case of the isotopic gauge transformation to counteract the dependence of $S$ on $x$, $y$, $z$ and $t$. (\dots) The field equations satisfied by the twelve independent components of the $B$ field, which we shall call
the $\bf b$ field, and their interaction with any field having an isotopic spin are essentially fixed,
in much the same way that the free electromagnetic field and its interactions with charged fields are essentially determined by the requirement of gauge invariance.}
%Quede aqui!

It was also clearly stated by Utiyama~\cite{Utiyama}:
{\em The form of the interactions between some well
known fields can be determined by postulating
invariance under a certain group of transformations.
For example, let us consider the electromagnetic interaction
of a charged field $Q(x)$, $Q^*(x)$. The electromagnetic
interaction appears in the Lagrangian through
the expressions
$\partial_\mu Q - i eA_\mu Q, \ \partial^\mu Q^*+ieA^\mu Q^*\ (1).$
The gauge invariance of this system is easily verified in
virtue of the combinations of $Q$, $Q^*$, and $A_\mu$ in (1), if
this system is invariant under the phase transformation
$Q\to e^{i\alpha} Q$, $Q^*\to Q^*e^{-i\alpha}$ $\alpha= {\rm const}\ (2)$
{\rm [because the combination $\partial_\mu Q - i eA_\mu Q$ also transforms as Q]. }
Reversing the argument, the combination (1) can be
uniquely introduced by the following line of reasoning.
In the first place, let us suppose that the Lagrangian
$L(Q,Q_{,\mu})$ is invariant under the constant phase transformation
(2). Let us replace this phase transformation
with the wider one (gauge transformation) having the
phase factor $\alpha(x)$ instead of the constant $\alpha$. In order to
make the Lagrangian still invariant under this wider
transformation it is necessary to introduce the electromagnetic
field through the combination (1). This
combination and the transformation character of $A_\mu$
under the gauge transformation can be uniquely determined
from the gauge invariance postulate of the
Lagrangian $L(Q,Q_{,\mu},A_\mu)$.}

\subsection{Conserved currents and equations of motion in the presence of gauge fields}

In the present context of an Abelian gauge theory, the free gauge field
contribution ${\cal L}_{U(1)}$ does not play any role in the expression of the conserved currents. This is due to the fact that
the bosons (photons here), which are responsible for the (electromagnetic) interaction, do not carry the corresponding (electric) charge. We shall see
later that this is no longer true in the non Abelian case.
Nevertheless, the conserved current gets an additional 
term due to the {presence} of the interaction of the matter field $\varphi$ with the gauge potentials~\cite{KaratasKowalski90}. 
Indeed, since any local phase transformation is now a symmetry of the
Lagrangian ${\cal L}_{\rm mat.}={\cal L}_0+{\cal L}_{\rm int.}$, this is also true for
global phase transformations and equation~(\ref{eq-continuity-j})
now becomes 
\begin{equation}
\delta({\cal L}_0+{\cal L}_{\rm int.})
=0=-\alpha(\partial_t\rho+\vec\nabla\cdot{\vecc J})
\label{eq-continuity-J}
\end{equation}
where the charge density $\rho({\vecc r},t)$ is unchanged, but the current
density acquires an additional term often called diamagnetic current density, 
${\vecc J}({\vecc r},t)=
\frac{-ie\hbar}{2m}(\varphi^*({\vecc r},t)\vec{\cal D}\varphi({\vecc r},t)
-\varphi({\vecc r},t)(\vec{\cal D}\varphi({\vecc r},t))^*)
={\vecc j}({\vecc r},t)
-(e^2/m){\vecc A}({\vecc r},t)\varphi^*({\vecc r},t)\varphi({\vecc r},t)$. 
In terms of 
this conserved current density, the interaction term is often 
%{improperly} (i.e. to {lowest powers} in the gauge fields)  
written  to {lowest powers} in the gauge fields as ${\cal L}_{\rm int.}
={\vecc J}\cdot{\vecc A}-\rho\phi+O(\vec A^2)$, where all fields depend in the 
general case
on the variables $({\vecc r},t)$.
%\revision{you say written improperly but then you don't explain why}

The total Langrangian density becomes
% \begin{widetext}
\begin{eqnarray}
{\cal L}_{\rm tot.}&=&{\cal L}_{\rm mat.}+{\cal L}_{U(1)}\nonumber\\
&&
=i\hbar\varphi^*(\partial_t+{\textstyle\frac{ie}\hbar}\phi)\varphi
-{\textstyle\frac{\hbar^2}{2m}}[(\vec\nabla-{\textstyle\frac{ie}\hbar}\vec A)\varphi]^*[(\vec\nabla-{\textstyle\frac{ie}\hbar}\vec A)\varphi]
-V\varphi^*\varphi\nonumber\\
&&\quad+{\scriptstyle\frac 12}\varepsilon_0(-\vec\nabla\phi-\partial_t{\vecc A})^2
-{\scriptstyle\frac 1{2\mu_0}}(\vec\nabla\times{\vecc A})^2
\end{eqnarray}
% \end{widetext}
The equations of motion follow
from Euler-Lagrange equations~(\ref{EL0}). For the matter field  equation of motion, we
perform the variation w.r.t. $\varphi^*({\vecc r},t)$,
\begin{eqnarray}
0&=&\frac\delta{\delta{\varphi^*}}\int 
{\cal L}_{\rm tot.} d^3r\ \! dt%\nonumber\\&=&
=
\frac{\partial {\cal L}_{\rm mat.}}{\partial{\varphi^*}}-\partial_j
\frac{\partial {\cal L}_{\rm mat.}}{\partial(\partial_j{\varphi^*})}
-\partial_t
\frac{\partial {\cal L}_{\rm mat.}}{\partial(\partial_t{\varphi^*})},\label{EL2}
\end{eqnarray} leading to 
\begin{equation}
i\hbar\partial_t\varphi = \frac{1}{2m}(-i\hbar\vec\nabla-e\vec A)^2\varphi
	+e\phi\varphi + V\varphi.
\end{equation}
Those for the gauge fields also
obey Euler-Lagrange equations, but follow from variations w.r.t. the gauge fields,
\begin{eqnarray}
0&=&\frac\delta{\delta{\modrm A}_k}\int 
{\cal L}_{\rm tot.} d^3r\ \! dt %\nonumber\\&=&
=
\frac{\partial {\cal L}_{\rm mat.}}{\partial{\modrm A}_k}-\partial_j
\frac{\partial {\cal L}_{U(1)}}{\partial(\partial_j{\modrm A}_k)}
-\partial_t
\frac{\partial {\cal L}_{U(1)}}{\partial(\partial_t{\modrm A}_k)},\label{EL1}\\
0&=&\frac\delta{\delta{\phi}}\int 
{\cal L}_{\rm tot.} d^3r\ \! dt%\nonumber\\&=&
=
\frac{\partial {\cal L}_{\rm mat.}}{\partial{\phi}}-\partial_j
\frac{\partial {\cal L}_{U(1)}}{\partial(\partial_j{\phi})}
%-\partial_t
%\frac{\partial {\cal L}_{\rm tot.}}{\partial(\partial_t{\phi})}
,\label{EL2}
\end{eqnarray}
and the very definition of the charge and current densities
$\rho=-\frac{\partial {\cal L}_{\rm mat.}}{\partial{\phi}}$,
$\vec{\modrm J}=\frac{\partial {\cal L}_{\rm mat.}}{\partial{\vecc A}}$.
The Maxwell equations follow
\begin{eqnarray}
&&\vec\nabla\times(\vec\nabla\times{\vecc A})
-\varepsilon_0\mu_0\partial_t(-\vec\nabla\phi-\partial_t{\vecc A})
=\mu_0{\vecc J},\label{Maxwell1}\\
&&\vec\nabla\cdot(-\vec\nabla\phi-\partial_t{\vecc A})=\frac\rho{\varepsilon_0},
\label{Maxwell2}
\end{eqnarray}
the two equations without sources being automatically satisfied with the 
definition of the physical fields ${\vecc E}$ and 
${\vecc B}$ in terms of the gauge fields. In the presence of other types
of interactions of the matter field, there can occur new contributions 
to the charge and current densities in Maxwell equations as we will see later.

The gauge field theory approach thus appears as a very powerful tool in order
to define properly the notion of conserved currents, the interaction of a
matter field with an external (gauge) field, the equations of motion of these
gauge fields, as well as some other subtle notions such as topological
invariants~\cite{WuYang1975}. This last notion is very useful for example in the description 
of the Aharonov-Bohm effect in which a charged particle moves in a pure
gauge field (i.e. a region of space where the physical field $\vecc B$ 
vanishes but the potential vector does not). %, ${\vecc A}=\vec\nabla\alpha$).
%\revision{I do not agree with this last statement, if it were true you could gauge it away, everywhere and you can't. Now pure gauge is generally thought as A being the gradient of a function. I would eliminate the last equation on the gradient}
In such a case, the gauge invariant phase of the wave function 
$\frac {ie}\hbar\oint{\vecc A}\ \!d{\vecc r}$ is responsible for the
shift, proportional to the magnetic flux enclosed by the path, 
of the interference pattern produced by any kind of {\em two-slit}
experiment~\cite{PeshkinTonomura}.

\section{Yang-Mills theory}\label{secIII}
Our program is now to follow the same lines of derivation {as previously outlined}, but in the
case of a SU(2) symmetry. This is done in many excellent 
textbooks~\cite{Ryder,Weinberg,PeskinSchroeder,Quigg,Aitchison,Ramond90}, 
so we will just motivate the theory and fix the notations that we will need in the
application {we have in mind} to condensed matter physics.
The U(1) gauge transformation introduced in the previous section was 
motivated by the fact that non relativistic (spinless) quantum mechanics deals
with wave functions which are complex scalar fields and that physics is
independent of the choice of global phase of this field. A complex scalar
field can also be seen as two real scalar fields (the real and imaginary 
parts), and the gauge transformation thus appears as a rotation in the
complex plane~\cite{Moriyasu80}.
We can imagine that a more elaborate theory would require the introduction of 
more sophisticated mathematical objects, e.g. spinors~\cite{Penrose} (or a set of complex
scalar fields). In the simplest case, we will assume that the  state of the 
particle is no longer described by a simple complex scalar field, but by a 
Pauli spinor consisting in two complex 
scalar fields usually written in the form of a ``vector'' 
$\psi=
\Bigl(
\begin{array}{c} 
%\hskip-2pt
\varphi_1%\hskip-2pt 
\\ [-1pt] 
%\hskip-2pt
\varphi_2%\hskip-2pt
\\[-0pt]
\end{array}
\Bigr)$ and we will call SU(2) {(special unitary group of $2\times 2$ matrices with determinant 1)}
 the corresponding gauge theory. {This name comes from the fact that the objects acting on {a} two component
 spinor are two by two matrices which can all be written in terms of four basic ones, the unit matrix and
 the Pauli matrices.}
Now, we obviously have a set of 4 real scalar fields, {but only 3 are
in fact} independent {because the wave function must be normalized as}
$\int d^3r\ \!\psi^\dagger\psi < +\infty$ where the adjoint spinor is the
``row vector'' $\psi^\dagger=(\varphi_1^*\ \varphi_2^*)$. 
In the ``internal'' space called
{\em isospin space} {(name given by Eugene Wigner)} where these
three components live, a gauge transformation may be seen now as a rotation
in a $3-$dimensional space. {Since the action of SU(2) operations rotate in 
three dimensional space and such rotations do not commute (order of rotations changes results) as in two dimensions, one says that the
theory is non Abelian}. {This contrasts with the Abelian U(1) theory where complex scalars commute under multiplication.}
Such a transformation, acting on the spinor, may be written as 
\begin{equation}
\psi({\vecc r},t)\to
\exp(ig{\boldsymbol\alpha}/\hbar)\psi({\vecc r},t).\label{transfSU2}
\end{equation} 
The constant $g$ plays the role of a new charge and all operators acting 
on the spinors are defined as $2\times 2$ matrices (as expected),
denoted here with bold font to emphasize {their matrix 
nature}, e.g. ${\boldsymbol\alpha}$ stands for the contraction  over the
repeated index $a$
${\boldsymbol\tau}^a\alpha^a=
{\boldsymbol\tau}^1\alpha^1+{\boldsymbol\tau}^2\alpha^2
+{\boldsymbol\tau}^3\alpha^3$ where the 
${\boldsymbol\tau}^a$ are the generators of SU(2) (i.e. half the Pauli 
matrices),
\begin{equation}
{\boldsymbol\tau}^1=\frac 12\begin{pmatrix}0&1\\ 1&0\end{pmatrix},\
{\boldsymbol\tau}^2=\frac 12\begin{pmatrix}0&-i\\ i&0\end{pmatrix},\ 
{\boldsymbol\tau}^3=\frac 12\begin{pmatrix}1&0\\ 0&-1\end{pmatrix}.
\end{equation}
The $\alpha^a$'s are the 
parameters of the transformation, they define the components of the rotation
vector in the three-dimensional isospin space.

The transformation (\ref{transfSU2}) is a symmetry %\revision{A symmetry is a invariance what do you mean here?} 
when the parameters 
$\alpha^a$'s are just constants, hence a conservation equation follows from 
Noether theorem, since, as quoted by R. Mills,
{\em ``The substance of the theorem, for our purposes, is that for every
symmetry of nature there is a corresponding conservation
law and for every conservation law there is a symmetry''}~\cite{Mills89}.
and its extension to local gauge symmetry is the strategy originally
proposed by Yang and Mills~\cite{YangMills54b}: {\em ``The conservation of 
isotopic
spin points to the existence of a fundamental invariance law
similar to the conservation of electric charge. In the latter
case, the electric charge serves as a source of electromagnetic
field. An important concept in this case is gauge invariance
which is closely connected with (1) the equation of motion
of the electromagnetic field, (2) the existence of a current
density, and (3) the possible interactions between a charged
field and the electromagnetic field. We have tried to generalize
this concept of gauge invariance to apply to isotopic spin
conservation''}. Isotopic spin or (strong) isospin had been introduced earlier by Heisenberg in 1933
as a new conserved physical quantity in order to explain the close similarity of behaviour of
protons and neutrons during strong interactions, e.g. the same cross section (apart from electromagnetic corrections) for interactions involving both types of particles, like
$p+\pi^-\to n+\pi^0$ or $n+\pi^+\to p+\pi^0$. The nucleon  was then considered as an isospin doublet.

The extension of the transformation (\ref{transfSU2})
to local parameters 
$\alpha^a(\vec r,t)$,  faces the problem that 
exactly like in the case of U(1) symmetry,
the space and time derivatives of the spinor do not obey the same 
transformation law (\ref{transfSU2}) as the spinor itself does. 
{As we mentioned before the simple gradient applied to the spinor
is not gauge invariant, and thus is not a valid building block for the Lagrangian.
The {alternative} is to come up with a recipe for a new derivative, the covariant
derivative, which after having acted on a spinor, behaves also like a spinor
(as one should expect, the derivative of a spinor is another spinor)}.

A space covariant derivative has to be defined as
\begin{equation}\vec{\nadermat}=\vec{\nabla}\nbOne
-{\textstyle\frac{ig}\hbar}\vec{\naAmat},\end{equation} or in component form
$\nadermat_k=\partial_k\nbOne
-{\textstyle\frac {ig}\hbar}
{\boldsymbol\tau}^a{\naA}_k^a$,
and a covariant 
time derivative is similarly introduced by, 
\begin{equation}{\nadermat}_t={\partial}_t\nbOne
+{\textstyle\frac{ig}\hbar}{\naphimat}={\partial}_t\nbOne
+{\textstyle\frac{ig}\hbar}{\boldsymbol\tau}^a{\naphi}^a.\end{equation}  
${\nbOne}$ is the $2\times 2$
identity matrix.
The non-Abelian gauge potentials $\naA^a_k$ and $\naphi^a$  
carry an isospin index, and they obey specific
transformation laws through local gauge
transformations.

\def\vertspace{
	\phantom{\begin{matrix}x\\x\end{matrix}}
	}
% \begin{widetext}
%\begin{center}
%\begin{table}
%\begin{sideways}
{%\footnotesize
\begin{equation}
\begin{array}{l}
\begin{array}{ccc}
\hline
\vertspace
U(1)&\hbox{Global gauge transformations}&SU(2)\nonumber\\ \hline
i\hbar\varphi^*\dot\varphi-\frac{\hbar^2}{2m}(\vec{{\nabla}}\varphi)^*
(\vec{{\nabla}}\varphi)%-V\varphi^*\varphi
&
{\rm free~particle~Lagrangian~density}~{\cal L}_0
&
i\hbar\psi^\dagger\dot\psi-\frac{\hbar^2}{2m}(\vec\nabla\nbOne\psi)^\dagger
(\vec\nabla\nbOne\psi)%-V\psi^\dagger\psi
\\
i\hbar\partial_t\varphi=-\frac{\hbar^2}{2m}\vec{{\nabla}}^2\varphi
&
{\rm free~particle~equation~of~motion}
&
i\hbar\partial_t\nbOne\psi=-\frac{\hbar^2}{2m}\vec{{\nabla}}^2\nbOne\psi
\\
\vertspace
\varphi\to\exp(ie\alpha/\hbar)\varphi
&
{\rm gauge~transf.}
&
\psi\to\exp(\frac i\hbar g%\vec{\boldsymbol\tau}\vec
{\boldsymbol\alpha})\psi
\\
\partial_t\rho+\vec\nabla\cdot{\vecc j}=0
&
{\rm continuity~equation}
&
\partial_t\rho^a+\vec\nabla\cdot{\vecc j}^a=0
\\
\vertspace
\rho=e\varphi^*\varphi
&
{\rm charge~density}
&
\rho^a=g\psi^\dagger{\boldsymbol\tau}^a\psi
\\
\vertspace
{\vecc j}=
\frac{-ie\hbar}{2m}(\varphi^*\vec\nabla\varphi
-\varphi\vec\nabla\varphi^*)
&
{\rm conserved~current}
&
{\vecc j}^a=
\frac{-ig\hbar}{2m}(\psi^\dagger{\boldsymbol\tau}^a(\vec\nabla\psi)
-(\vec\nabla\psi)^\dagger{\boldsymbol\tau}^a\psi)\\
\hline
\end{array}\\
\vertspace\hbox{{\bf Table 1:} Comparison between U(1) and SU(2) global gauge invariance.}
\end{array}
\end{equation}
} % footnotesize
%\end{sideways}
% \end{widetext}

Once the non-Abelian gauge potentials have been defined, we can form 
a field tensor and a gauge covariant Lagrangian density in order to allow 
for a simple derivation
of the conserved current density following from Noether theorem, 
and essentially follow similar lines of derivation than in the previous
section. In order not to repeat all calculations, we present the analogies 
between U(1) and SU(2) theories in
two tables. The properties associated to global gauge 
invariance are listed in the first table, 
while the consequences of local gauge invariance follow in the next table.

Let us briefly comment upon the results listed in
these tables. In Table~1 there is a very clear one to one correspondence between the quantities defined in the two gauge theories.
When one invokes global gauge invariance in the SU(2) case, 
%angular momentum \revision{where is angular momentum, you have not said, rho is the angular momentum density in SU(2) correct?}
isospin is inert fixed in a certain direction and everything operates as if charge is the only physical
quantity. This is consistent with the interpretation of the Schr\"odinger equation not as an equation of motion for spinless particles, but
for ``two-component Pauli particles'' in a spin eigenstate~\cite{HestenesGurtler75}.
There are nevertheless important differences 
between the Abelian and the non-Abelian theories. First, the fact that the
gauge potentials carry internal indices (this is the origin of the non-Abelian
character), implies that the same indices are carried by the conserved charges,
as well as by all other fields such as $\vec\naE^a$ and $\vec\naB^a$, the
non-Abelian {analogues} to the electric and magnetic fields defined in Table~2.
Another essential difference is due to the fact that the gauge fields also
carry the ``charge'' of the interaction. This has two main consequences (see
Table 2):
the conserved charge density and current density have, together with
the matter part, a purely {\em radiative} {(i.e.  involving only the gauge fields)}
contribution and the equations of motion for the gauge fields are non-linear.

% \begin{widetext}
%\begin{sideways}
{\footnotesize
\begin{equation}
\begin{array}{l}
\begin{array}{ccc}
\hline
\vertspace
U(1)
&\begin{array}{c}\hbox{Local gauge}\\ \vspace{-5mm}\\ 
	\hbox{transformations}\end{array}
&SU(2)\nonumber\\ \hline\\ \vspace{-8mm}
\varphi\to\exp(ie\alpha(\vecc r,t)/\hbar)\varphi=G\varphi
&
{\rm local~gauge~transf.}
&
\psi\to\exp(\frac i\hbar g%\vec{\boldsymbol\tau}\vec
{\boldsymbol\alpha}(\vecc r,t))\psi=\naGmat\psi
\\ \vspace{8mm}
\\
\left.\begin{array}{c}
\vec{\cal D}=\vec\nabla-\frac{ie}{\hbar}{\vecc A}\\
{\cal D}_t=\partial_t+\frac{ie}{\hbar}\phi
\end{array}\right\}
&
{\rm cov.~derivatives}\vertspace
&
\left\{\begin{array}{c}
\vec{\nadermat}=\vec{\nabla}\nbOne
-\frac{ig}\hbar\vec{\naAmat}\\
{\nadermat}_t={\partial}_t\nbOne
+\frac{ig}\hbar{\naphimat}
\end{array}\right.
\\ \vspace{0mm}\\
\left.
\begin{array}{l}
{\vecc A}\to {\vecc A}+\vec\nabla\alpha\\
\phi\to\phi-\partial_t\alpha
\end{array}\right\}
&
\begin{array}{c}
{\rm gauge~transf.}\\ \vspace{-5mm}\\
{\rm of~potentials}
\end{array}&
\left\{
\begin{array}{l}
\vec{\naAmat}\to \vec{\naAmat}+ 
\vec\nabla{\boldsymbol\alpha}
+\frac{ig}{\hbar}
[{\boldsymbol\alpha},\vec{\naAmat}]
\\
{\naphimat}\to {\naphimat}
-
\partial_t{\boldsymbol\alpha}
+\frac{ig}{\hbar}
[{\boldsymbol\alpha},{\naphimat}]
\end{array}\right.
\\  \vspace{0mm}\\
\left.\begin{array}{l}
{\modrm E}_k=\frac{\hbar}{ie}[{\cal D}_t,{\cal D}_k]
\\
{\modrm B}_k=-\frac{\hbar}{2ie}\epsilon_{ijk}[{\cal D}_i,{\cal D}_j]
\end{array}\right\}
&
{\rm field~strengths}
&
\left\{
\begin{array}{l}
{\naEmat}_k=\frac{\hbar}{ig}[\nadermat_t,\nadermat_k]
\\
{\naBmat}_k=-\frac{\hbar}{2ig}\epsilon_{ijk}[\nadermat_i,\nadermat_j]
\end{array}\right.
\\  \vspace{0mm}\\
\begin{array}{c}
i\hbar\varphi^*{\cal D}_t\varphi
-\frac{\hbar^2}{2m}(\vec{{\cal D}}\varphi)^*
(\vec{{\cal D}}\varphi)
\\
+\frac{\varepsilon_0}2(|\vec{\modrm E}|^2-c^2|\vec{\modrm B}|^2)
\end{array}
&
{\rm Lagrangian}~{\cal L}_{\rm tot.}
&
\begin{array}{c}
i\hbar\psi^\dagger{\nadermat}_t\psi-\frac{\hbar^2}{2m}(\vec{\nadermat}\psi)^\dagger
(\vec{\nadermat}\psi)
\\
+\frac{g}{2c}(\vec{\naE}^a\cdot\vec{\naE}^a-c^2\vec{\naB}^a\cdot\vec{\naB}^a)
\end{array}
\\ \vspace{0mm}\\
\rho=e\varphi^*\varphi
&
{\rm cov.~charge~density}
&
\rho^a_{\rm mat.}+\rho^a_{\rm rad.}=g\psi^\dagger {\boldsymbol\tau}^a\psi
-\frac{g}{c}\epsilon_{abc}\vec{\naA}^b\cdot\vec{\naE}^c
\\ \vspace{0mm}\\
\begin{array}{c}
\vec{\modrm J}=
\frac{-ie\hbar}{2m}(\varphi^*(\vec\nabla\varphi)
-\varphi(\vec\nabla\varphi)^*)\\
-\frac{e^2}m\vec{\modrm A}\varphi^*\varphi
\end{array}
&
\begin{array}{c}
{\rm cov.~conserved~}\\ \vspace{-5mm}\\
{\rm current}
\end{array}
&
\begin{array}{c}
\vec{\modrm J}^a+\vec{\ \!\cal J}^a=
\frac{-ig\hbar}{2m}(\psi^\dagger {\boldsymbol\tau}^a
(\vec\nabla\psi)
-(\vec\nabla\psi)^\dagger {\boldsymbol\tau}^a\psi)
\\
-\frac{g^2}{4m}\vec{\naA}^a\psi^\dagger\psi
-\frac{g}{c}\epsilon_{abc}(\naphi^b\vec\naE^c+{\textstyle \frac 12}c^2
\vec\naA^b\times\vec\naB^c)
\end{array}
\\
\vspace{0mm}\\
i\hbar\partial_t\varphi = \frac{1}{2m}(-i\hbar\vec\nabla-e\vec A)^2\varphi
	+e\phi\varphi 
&
\begin{array}{c}
{\rm matter~field}\\  \vspace{-5mm}\\
{\rm equation~of~motion}
\end{array}
&
i\hbar\partial_t\nbOne\psi=\frac{1}{2m}(-i\hbar\vec{{\nabla}}\nbOne
-g\vec\naAmat)^2\psi	+g\naphimat\psi 
\\ 
\vspace{0mm}\\
\left.\begin{array}{l}
\vec\nabla\times{\vecc B}
-c^{-2}\partial_t{\vecc E}
=\mu_0{\vecc J}\\
\vec\nabla\cdot{\vecc E}=\frac\rho{\varepsilon_0}
\end{array}
\right\}
&
\begin{array}{c}
{\rm gauge~fields}\\  \vspace{-5mm}\\
{\rm equations~of~motion}
\end{array}
&
\left\{
\begin{array}{l}
\vec\nabla\times{\vecc\naB}^a
-c^{-2}\partial_t{\vecc \naE}^a
=\frac{1}{g c}({\vecc J}^a+\vec{\ \!\cal J}^a)\\
\vec\nabla\cdot{\vec\naE}^a=\frac cg(\rho^a_{\rm mat.}+\rho^a_{\rm rad.})
\end{array}\right.
\\ \vspace{-1mm}\\
\hline
\end{array}\\
\vertspace\hbox{{\bf Table 2:} Comparison between U(1) and SU(2) local gauge invariance.}
\end{array}
\end{equation}
} % footnotesize
%\end{sideways}
%\end{table}
%\end{center}
% \end{widetext}

Another subtle difference between the Abelian and the non Abelian case resides in which quantities are
gauge invariant, which are gauge covariant, and which are simply gauge dependent. One is used to the case of the electromagnetic interaction
for which the charge density and the total current density are gauge invariant (even though in the latter case the paramagnetic and diamagnetic
contributions are separetaly not gauge invariant, their sum is). The electric and magnetic fields (or the Faraday
tensor) are similarly gauge invariant and finally only the wave function is gauge covariant $\varphi'=G\varphi$. The covariant derivative
operator ${\cal D}_\mu$ 
introduced to satisfy the same covariance law $({\cal D}_\mu\varphi)'\equiv {\cal D}'_\mu\varphi'=G{\cal D}_\mu\varphi$ then tranforms as a vector
${\cal D}'_\mu=G{\cal D}_\mu G^{-1}$ and this implies that the gauge potential, although a $4-$vector, {has} a particular non-covariant
gauge transformation $A_\mu'=G A_\mu G^{-1}-\frac\hbar{ie}(\partial_\mu G)G^{-1} $.
In the non Abelian situation, the same consideration {applies} for the spinor $\psi'=\naGmat\psi$ and the covariant derivative
operator, ${\nadermat}'_\mu=\naGmat{\nadermat}_\mu \naGmat^{-1}$, and this requires the gauge potential to obey a similar non-covariant
gauge transformation $\naAmat_\mu'=\naGmat \naAmat_\mu \naGmat^{-1}-\frac\hbar{ig}(\partial_\mu \naGmat)\naGmat^{-1} $. But now the so-called {\em curvature} tensor $\naFmat_{\mu\nu}$ transforms like a vector, $\naFmat'_{\mu\nu}=\naGmat{\naFmat}_\mu \naGmat^{-1}$, i.e. is subject to a rotation in the isospin space, hence  the SU(2) magnetic and electric fields are not gauge invariant and are then unphysical in a gauge symmetric theory (for a discussion
on the {\it meaning of ``physical meaning''}, see e.g. Ref.~\cite{Giuliani}). 
The non Abelian charge density and current density themselves are {\em not} gauge invariant, and under a gauge transformation,
their matter and radiation content changes. Only the total non Abelian charge $Q^a=\int d^3r\ \!\rho^a(\vec r,t)$ is gauge invariant (and constant in time)
provided that the volume of integration extends over large enough distances for the current to vanish.

\section{Yang-Mills theory for spin-orbit interaction}\label{secIV}
\subsection{Preamble - the Zeeman interaction}
Electrons are particles which carry not only an electric charge, but also
a spin $\frac 12$. In non-relativistic
Quantum Mechanics, they obey the Pauli equation which is a generalization of
the Schr\"odinger equation with an essential innovation: the existence of the 
spin, i.e. a new degree of freedom, coupled to space-time degrees of freedom 
in the Hamiltonian through interactions such as {the} spin-orbit interaction {and the}
Zeeman interaction. In  the presence of such interactions, spin is known not to be 
conserved, but spin {carries} angular momentum and the total angular 
momentum is conserved. This is at the origin e.g. of the Einstein - de Haas
experiment that will be discussed later. 

Although these statements are well understood physically, a clear formulation {is appealing
following} the lines of derivation proposed in the quotation of Yang and 
Mills~\cite{YangMills54b} in section~\ref{secIII}. 
{Here the starting point is given by the known interaction terms present in 
the Pauli equation, from which a consistent gauge theory is built. {Such a formulation
leads, as before, to a conservation law. Then, once the gauge transformations
are extended to local changes, it leads to the equations of motion for the gauge fields.}

In this introductory section, we will focus attention 
on the consequences of the Zeeman interaction between the magnetic 
moment associated to the 
spin of the electron $\frac {e}{m}\vec{\bf s}$ (with Land\'e factor $g_e=2$
and $\vec{\bf s}=\frac 12\hbar\vec{\boldsymbol\sigma}$)
and an external magnetic field $\vec B$.
The state of the particle is described by 
a  Pauli spinor
$\psi=
\Bigl(
\begin{array}{c} 
%\hskip-2pt
\varphi_\uparrow%\hskip-2pt 
\\ [-1pt] 
%\hskip-2pt
\varphi_\downarrow%\hskip-2pt
\\[-0pt]
\end{array}
\Bigr)$, and the Hamiltonian reads as 
$\mat H=
\frac 1{2m}(\vecc{p}-e\vecc{A})^2\nbOne
+ V\nbOne-
\frac{e}{m}\vec{{\bf s}}\cdot{\vecc B}%}_{{\tt Zeeman~term}}
$. Following standard textbooks~\cite{Landau,Davydov}
 one can build a continuity equation where the charge
density and the charge current density are defined according to 
\begin{eqnarray}
\rho({\vecc r},t)&=&
e\psi^\dagger({\vecc r},t)\psi({\vecc r},t),\\%$ and$
{\vecc J}({\vecc r},t)&=&
\frac{-ie\hbar}{2m}(\psi^\dagger({\vecc r},t)[\vec\nabla\psi({\vecc r},t)]
-[\vec\nabla\psi^\dagger({\vecc r},t)]\psi({\vecc r},t))\nonumber\\
&&-\frac{e^2}{m}\vec A
\psi^\dagger\psi+\frac em
\vec\nabla\times(\psi^\dagger\vec{\bf s}\psi).\end{eqnarray}
As we had already anticipated, when a new interaction (here the Zeeman 
interaction) appears in the problem, 
the conserved current gets modified. 
This is a property that we have already
seen at play when the minimal coupling to the external electromagnetic field
was added to the free particle, {where it led 
to the appearance of a diamagnetic contribution to
the Noether conserved current density.}
Here, the rotor of the magnetization
associated to the electron density produces an additional charge 
current density. Note that this last term may be forgotten during the 
standard derivation using the wave equation and its
complex conjugate, since the divergence of a rotor vanishes, and some care must
be taken to establish the full current 
density~\cite{Hestenes79,Nowakowski,ShikakhwaEtAl}.
Nevertheless, this addititonal term, sometimes called ``spin term''~\cite{Nowakowski},
is compulsory in order to obtain a 
conservation equation.
{It is thus also instructive to contemplate} 
the spin density (e.g. in the direction of
the external field $\vec B=B\vec e_z$), which may be naturally 
defined by 
$s^z({\vecc r},t)=\psi^\dagger{\bf s}^z\psi$.
The $z-$component ${\bf s}^z$ commutes with
the Hamiltonian. This is a conserved quantity, and the corresponding
continuity equation takes the form ${\partial_t s^z}+\vec\nabla\cdot\vec
J^z=0$ with $\vec J^z=\frac{-i\hbar}{2m}(\psi^\dagger{\bf s}^z\vec\nabla\psi
-(\vec\nabla\psi^\dagger){\bf s}^z\psi)$ the spin current density.
This is a tensorial quantity, since it has a vector character (the
current propagates in space) and depends also on the spin orientation. The 
conservation equation also follows from the use of Ehrenfest theorem, 
$\frac{d s^z}{dt}=(i\hbar)^{-1}\psi^\dagger[{\bf s}^z,\mat H]\psi +\psi^\dagger{\partial_t{\bf s}^z}\psi=0$.
The other
components of the spin however 
do not commute with the Hamiltonian and in this case the continuity
equation gets additional terms. Altogether, one can write
${\partial_t s^a}+\vec\nabla\cdot({s^a\vec v})=\frac em(\vec s\times\vec B)_a$,
where the RHS represents the
torque exerted on the spin. 

{The spin} is just a contribution to the total angular momentum and this latter quantity is of course conserved.
Our discussion thus suggests that there should be a way to write a conservation equation where spin components would appear explicitly,
together with other sources of angular momentum. The ``spin density'' components $s^a({\vecc r},t)=\psi^\dagger{\bf s}^a\psi$ {are, in a sense equivalent} to the charge density, except that they carry the spin index $a$. Deriving such a conservation equation
is the purpose of the next sections.

\subsection{Pauli equation and spin-orbit interaction}
We are now in a position to discuss completely the role of the spin-dependent
relativistic corrections to the Schr\"odinger equation.
The Pauli equation follows from the non relativistic limit of the Dirac
equation~\cite{BjorkenDrell},
\begin{equation}
i\hbar\partial_t\Psi=\big[c\vec\alpha\cdot(\vec p-e\vec A)+\beta mc^2+V\big]
\Psi,
\end{equation}
where $\vec \alpha$ and $\beta$ are the $4\times 4$ 
Dirac matrices and $\Psi$ is a 
$4-$component Dirac spinor. We let $\Psi=\Bigl(
\begin{array}{c} 
%\hskip-2pt
\psi%\hskip-2pt 
\\ [-1pt] 
%\hskip-2pt
\xi%\hskip-2pt
\\[-0pt]
\end{array}
\Bigr)e^{-imc^2t/\hbar}$ to remove the rest energy of the electron. 
$\psi$ and $\xi$ both have two components.
In the 
non relativistic limit, $\xi$ is a very small component
$\xi\simeq \frac{1}{mc^2-V}c\vec{{\boldsymbol\sigma}}\cdot(\vec p-e\vec A)\psi$, and $\psi$
satisfies
\begin{equation}
i\hbar\partial_t\nbOne\psi=\mat H\psi.
\end{equation}
Here, $\vec{{\boldsymbol\sigma}}$ is the vector of Pauli matrices
$\vec{{\boldsymbol\sigma}}={\boldsymbol\sigma^1}{\vecc e}_x
+{\boldsymbol\sigma^2}{\vecc e}_y+{\boldsymbol\sigma^3}{\vecc e}_z
={\boldsymbol\sigma^a}{\vecc e}_a$. Forgetting about the relativistic correction to the kinetic energy, 
the Pauli Hamiltonian
is given by
% \begin{widetext}
\begin{eqnarray}
\hskip-20mm
\mat H=\underbrace{\frac 1{2m}(\vecc{p}-e\vecc{A})^2\nbOne}_{{\tt
Non~Relativistic~KE}}+ V\nbOne
&&
\nonumber\\ &&\hskip-30mm
-\underbrace{\frac{e\hbar}{2m}\vec{{\boldsymbol\sigma}}\cdot
{\vecc B}}_{{\tt Zeeman~term}}
-\underbrace{
\frac{e\hbar}{4m^2c^2}\vec\bsigma\cdot
( {\vecc E}\times ({\vecc p}-e{\vecc A}))
-\frac{ie\hbar^2}{8m^2c^2}\vec\bsigma\cdot\vec\nabla\times{\vecc E}
}_{{\tt SO~interaction}}
+\underbrace{\frac{e\hbar^2}{8m^2c^2}\vec\nabla^2\phi}_{{\tt Darwin~term}}
.\label{eq-Hamtot}
\end{eqnarray}
% \end{widetext}
  In the following, we will neglect the
Darwin interaction which, like the relativistic correction to the 
kinetic energy, is spin-independent. The ``substrate'' potential $V$ will
be denoted as $e\phi$, defining the scalar gauge potential. Let us note that the 
SO interaction is written in its more general form here, including the term proportional to the rotor of the electric field~\cite{BjorkenDrell}.

\subsection{Gauge theory of the spin-orbit interaction}
The fact that the SO interaction is linear in $\vecc p$ enables one
to gather it together with the Non Relativistic KE in a single squared term,
$\frac 1{2m}[(\vec p-e\vec A)\nbOne-g\vec\naAmat]^2$ 
up to a correction
quadratic in the spin operator. Since $\frac 12{\boldsymbol\sigma}^a$ is
the generator of SU(2) symmetry (denoted as ${\boldsymbol\tau}^a$ in tables 1 and 2), we set $\vec\naAmat
=\frac 12{\boldsymbol\sigma}^a\vec \naA^a$, with $\vec \naA^a$ three ordinary vectors.
Expanding the cross products in terms of antisymmetric tensors, 
the SO interaction in Eq.~(\ref{eq-Hamtot}) may now be written as
\begin{eqnarray}
\mat H_{{\tt SO}}&=&-\frac{e\hbar}{4m^2c^2}[\bsigma^a\epsilon_{ajk}
E_jp_k+\frac{i\hbar}2\bsigma^a\epsilon_{ajk}\partial_jE_k]
%\nonumber\\
\end{eqnarray}
and identified to the relevant contribution of the cross term in the Non Relativistic KE, 
\begin{eqnarray}
\mat H_{{\tt SO}}&=&-\frac g{4m}\bsigma^a(p_k\naA^a_k+\naA^a_kp_k)%\nonumber\\&=&
=\frac {i\hbar^2}{4m}\bsigma^a(2\naA^a_k\partial_k+(\partial_k\naA^a_k)),
\end{eqnarray}
 provided that the {\em new} ``SO'' charge,  here denoted as $g$,  is 
 identified
to $g=\hbar$. 
It is thus natural to define a new gauge vector potential, depending on both
space-time variables {\em and} spin degrees of freedom, 
$g\vec{\naAmat} =\frac{e\hbar}{4mc^2}
\vec{{\boldsymbol\sigma}}\times{\vecc E}
$~\cite{BercheEtAl12} or in components
\begin{equation}
{\scriptstyle\frac 12}g
{\boldsymbol\sigma}^a\vec{\naA}^a
\equiv{\scriptstyle\frac 12}g
{\boldsymbol\sigma}^a{\naA}^a_k\vec{\modrm e}_k=\frac{e\hbar}{4mc^2}
\epsilon_{ajk}{{\boldsymbol\sigma}^a}{\modrm E}_j\vec{\modrm e}_k.
\end{equation}
The same identification is obtained if we include the electromagnetic vector
potential.
We see that the gauge vector 
components  ${\naA}^a_k$ carry {\em two} indices. The index $k$ here refers to a purely space
index while the index $a$ is a spin index
which also has a spatial meaning 
due to the fact that spin and space variables are linked
in Quantum Mechanics (the index $a$ would be
a purely internal isospin variable in the context of Yang-Mills
theories of fundamental interactions). In order to avoid any confusion between
the two indices, we will always denote as upperscript the spin indices (the
position of the indices makes no difference, since we use a Euclidean metric
for this non relativistic theory). Repeated indices are summed up, {independently}
of their up or down position.
The  SO
interaction is now essentially included in the KE through %$\frac 1{2m}(\vec{\modrm p}\ \nbOne-e\vec{\modrm A}\ \!\nbOne-g\vec{\naAmat})^2$:
% \begin{widetext}
\begin{eqnarray}
\hskip-20mm
{\tt Non~Relativistic~KE\ }+{\ \tt SO~Interaction}= &&\nonumber\\
\hskip-0mm
\frac1{2m}\bigl(\vec{\modrm p}\ \nbOne-e\vec{\modrm A}\ \!\nbOne
-\frac{e\hbar}{4mc^2}
\vec{{\boldsymbol\sigma}}\times{\vecc E}\bigr)^2
-\frac{e^2\hbar^2}{32m^3c^4}|\vec{{\boldsymbol\sigma}}\times{\vecc E}|^2.
\label{eqKE+SO}
\end{eqnarray}
% \end{widetext}

Let us point out the important fact that the gauge vector potential has an
inherent matrix structure, hence the underlying gauge theory is  a
non-Abelian one, called SU(2) due to the presence of the Pauli matrices,
exactly like what was {briefly} presented in section~\ref{secIII}. 
The addititional term in equation~(\ref{eqKE+SO}) 
has the structure of a gauge symmetry breaking term~\cite{Moriyasu80b} i.e. {it
changes when we perform a gauge transformation. This, we now know how to do
for this theory, and its effect will
appear in a later discussion.} 

In the case of the SO$+$Zeeman interactions of equation~(\ref{eq-Hamtot}), 
one has to build a full
U(1)$\times$SU(2) theory. This means that we contemplate the space of complex scalar and their products and the space of
$2\times 2$ matrices and their product. Both these spaces being independent. To take into account simultaneously the fact that
the electric and magnetic fields are minimally coupled to the charge {\em and}
to the spin degrees of freedom,  we define the covariant derivatives
\begin{equation}
\begin{array}{c}
\vec{\nadermat}=\vec{\nabla}\nbOne-\frac{ie}\hbar\vec{\modrm A}\nbOne
-
i\vec{\naAmat},\\
{\nadermat}_t={\partial}_t\nbOne+\frac{ie}\hbar{\phi}\nbOne
+
i{\naphimat}.
\end{array}
\end{equation}
 where $\vec A$ is the electromagnetic vector potential 
and ${\naA}^a_k=\frac{e}{2mc^2}\epsilon_{ajk}{\modrm E}_j$,
while
$\phi$ is
the usual electromagnetic scalar potential, and $\naphi^a=
-\frac{e}{m}{\modrm B}^a$. 
Note that we have used the simplification which occurs due to the fact that the non-Abelian charge is $g=\hbar$.

Under an infinitesimal local gauge transformation 
$\naGmat=\exp(i\frac 12{\boldsymbol\sigma}^a\alpha^a)\simeq \nbOne
+i\frac 12{\boldsymbol\sigma}^a\alpha^a$, the non Abelian gauge fields
should transform according to 
\begin{eqnarray}
\naphimat\to\naphimat'&=&
\naGmat\naphimat\naGmat^{-1}-\frac 1i(\partial_t\naGmat)\naGmat^{-1}
%\nonumber\\ [-6pt]&=&
={\textstyle \frac 12}{\boldsymbol\sigma}^a(\naphi^a-\epsilon_{abc}\alpha^b\naphi^c-\partial_t\alpha^a)\\
\naAmat\to\naAmat'&=&\naGmat\naAmat\naGmat^{-1}+\frac 1i(\vec\nabla\naGmat)\naGmat^{-1}%\nonumber\\[-6pt]&=&
={\textstyle \frac 12}{\boldsymbol\sigma}^a(\vec\naA^a-\epsilon_{abc}\alpha^b\vec\naA^c+\vec\nabla\alpha^a).\end{eqnarray}
{The alert student should note that {\it in this theory of the spin-orbit interaction}} there 
is no gauge freedom associated to SU(2) transformations {(actually a small subset survives\cite{Medina})} since the non-Abelian gauge fields are defined in terms of the physical Maxwell fields~\cite{BercheEtAl12}. We will come back later to this point.
Another peculiarity is due to the appearance in equation~(\ref{eqKE+SO}) of 
the so-called {\em gauge symmetry breaking} term (GSB) (the last term in the
equation) {that interferes with gauge symmetry}. The full Lagrangian hence becomes ${\cal L}_{\rm tot.}=
{\cal L}_0+{\cal L}_{\rm int.}+{\cal L}_{\rm GSB}+{\cal L}_{U(1)}
+{\cal L}_{SU(2)}$, where ${\cal L}_{\rm mat.}
\equiv{\cal L}_0+{\cal L}_{\rm int.}+{\cal L}_{\rm GSB}$ 
(${\cal L}_{\rm mat.}$ contains all terms involving the matter field $\psi$).
We also have to add the free field contribution ${\cal L}_{SU(2)}$ 
(called Yang-Mills 
Lagrangian density), which only depends on the non-Abelian fields. 
This term is required by gauge symmetry and 
Lorentz invariance (for the theory to be automatically extended to a relativistic context).
% \begin{widetext}
\begin{eqnarray}
\hskip-10mm
{\cal L}_{\rm tot.}=
\underbrace{
i\hbar\psi^\dagger{\nadermat}_t\psi
-\frac{\hbar^2}{2m}({\nadermat}_k\psi)^\dagger
({\nadermat}_k\psi)
}_{{\cal L}_{0}+{\cal L}_{\rm int.}}
+
\underbrace{
\frac{\hbar^2}{2m}\psi^\dagger\naAmat_k\naAmat_k\psi
}_{{\cal L}_{\rm GSB}}
&&\nonumber\\ 
&&\hskip-50mm
+
\underbrace{
\frac{\varepsilon_0}2({\modrm E}_k{\modrm E}_k-c^2{\modrm B}_k{\modrm B}_k)
}_{{\cal L}_{U(1)}}
+
\underbrace{
\frac{\hbar}{2c}({\naE}_k^a{\naE}_k^a-c^2{\naB}_k^a{\naB}_k^a)
}_{{\cal L}_{SU(2)}}
,
\end{eqnarray}
% \end{widetext}
with 
\begin{eqnarray}
\vec\naE^a&=&-\partial_t\vec\naA^a-\vec\nabla\naphi^a+\epsilon_{abc}
\naphi^b\vec\naA^c,\label{eqnaE}\\
\vec\naB^a&=&\vec\nabla\times\vec\naA^a
+{\scriptstyle\frac 12}\epsilon_{abc}\vec\naA^b\times\vec\naA^c,\label{eqnaB}
\end{eqnarray}
following from the definitions 
$\naEmat_k=\frac 1i[\partial_t\nbOne+i\naphimat,\partial_k\nbOne-i\naAmat_k]$
and 
$\naBmat_k=-\frac 1{2i}\epsilon_{ijk}[\partial_i\nbOne-i\naAmat_i,
\partial_j\nbOne-i\naAmat_j]$ {given in Table 2}.
The dimensions of the fields are consistent with the coefficients
appearing in the covariant derivatives, $[\naphi^a]=\di T^{-1}$, 
$[\naA^a_k]=\di L^{-1}$, $[\naE^a_k]=\di L^{-1}\di T^{-1}$
and $[\naB^a_k]=\di L^{-2}$, {where $\di T$ and $\di L$ are time and space dimensions}. The fact that the
pure non-Abelian field contribution ${\cal L}_{SU(2)}$ is gauge invariant is
not {\em a priori} obvious due to the gauge co-variance  of the 
physical fields in the non-Abelian situation 
$\naEmat_k\to \naGmat\naEmat_k\naGmat^{-1}
=\naEmat_k+i\frac 12\alpha^a[{\boldsymbol\sigma}^a,\naEmat_k]$ and the similar
transformation law for
$\naBmat_k$.
Nevertheless, ${\cal L}_{SU(2)}$ being proportional to 
${\rm Tr}\ \!\naFmat_{\mu\nu}\naFmat^{\mu\nu}\to
{\rm Tr}\ \!\naGmat\naFmat_{\mu\nu}\naFmat^{\mu\nu}\naGmat^{-1}$, it is 
obviously gauge invariant thanks to the invariance properties of the trace under
{cyclic reordering of the matrices}. There remains the question of the physical origin of this 
term in the present situation. As argued by Tokatly~\cite{Tokatly}, it has
the correct form to account for the first SO corrections to the energy. In any case
it is demanded by symmetry so it should be out there!

\subsection{Conserved currents}
Let us stress on the fact that, contrary to the results presented in 
Table 2 in the case of a SU(2) {\it gauge symmetric theory} (Lagrangian invariant under SU(2) gauge transformations), 
the so called dia-color  contribution 
$-\frac{\hbar^2}{4m}\vec\naA^a\psi^\dagger\psi$ to the current density
is exactly compensated in the present situation 
by the GSB contribution. The term dia-color was coined in this context by Tokatly and refers to the analogue of the diamagnetic
current density, but for the non Abelian SO interaction~\cite{Tokatly}.
The matter current
${\modrm J}_k^a\equiv\frac{ \partial{\cal L}_{\rm mat.} }{ \partial\naA_k^a}$
finally simplifies from the expression in Table 2 to
\begin{equation}
{\modrm J}_k^a=\frac{-i\hbar^2}{2m}\bigl(\psi^\dagger{\scriptstyle \frac 12}
{\boldsymbol\sigma}^a
(\partial_k\psi)
-(\partial_k\psi)^\dagger{\scriptstyle \frac 12}
{\boldsymbol\sigma}^a\psi\bigr)\label{jtot},
\end{equation}
while the radiation current 
${\cal J}_k^a\equiv\frac{ \partial{\cal L}_{SU(2)}}{\partial\naA_k^a}$ 
is the same as indicated in the 
table,
\begin{equation}
{\cal J}_k^a=-{\textstyle \frac{\hbar}{c}}\epsilon_{abc}(\naphi^b\naE^c_k
+{\textstyle \frac 12}c^2
\epsilon_{ijk}\naA^b_i\naB^c_j).\label{eqn27}
\end{equation}
The GSB term has no effect on the matter and radiation
spin densities which remain those of the table, 
\begin{equation}
\rhospin^a_{\rm mat.}\equiv-\frac{ \partial{\cal L}_{\rm mat.} }
{ \partial\naphi^a}
=\psi^\dagger{\scriptstyle \frac 12}\hbar{\boldsymbol\sigma}^a\psi
\end{equation} 
and 
\begin{equation}\rhospin^a_{\rm rad.}
\equiv-\frac{ \partial{\cal L}_{SU(2)} }{ \partial\naphi^a}=
-{\textstyle \frac{\hbar}{c}}\epsilon_{abc}{\naA}^b_k{\naE}^c_k.
\end{equation}
The spin-dependent part in the interaction term  ${\cal L}_{\rm int.}$ 
is thus exactly $J_k^a\naA_k^a-\rhospin_{\rm mat.}^a\naphi^a$ (with no quadratic
contribution this time). A very simple, but important consequence is that there is no spin current contribution proportional to the non Abelian gauge vector,
i.e. no current transverse to the ordinary electric field, and the spin Hall conductivity vanishes automatically in the GSB scenario~\cite{Medina}.

\subsection{Equations of motion}
In order to obtain the equations of motion for the gauge fields, 
we also have to take into account 
the last peculiarity we will consider of the present problem, i.e. the fact that the U(1) and
the SU(2) gauge 
fields are not independent. {There are two approaches we can follow to contemplate this dependence within} the variational formulation: one either includes constraints using {the Lagrange multipliers~\cite{BercheEtAl12} method, which allows treating} the different gauge 
fields as independent,
% \begin{widetext}
\begin{eqnarray}
&S_{\rm constr.}[\psi,\phi,A_k,\naphi^a,\naA^a_k]
&=S[\psi,\phi,A_k,\naphi^a,\naA^a_k]\nonumber\\
&&\hskip-15mm -\int dt\ \! d^3r\left[\mu^a(\naphi^a+\frac em B^a)
+\lambda_i^a(\naA_i^a-\frac e{2mc^2}\epsilon_{aji}E_j)\right],\nonumber\\
&S[\psi,\phi,A_k,\naphi^a,\naA^a_k]
&=
\int dt\ \! d^3r
{\cal L}_{\rm tot.}(\psi,\phi,A_k,\naphi^a,\naA^a_k),\ 
\end{eqnarray}
% \end{widetext}
{or one has to include in the equations of motion an account of the second order derivatives in the U(1) gauge fields}, e.g.
%(\revision{why should in principle this latter approach work?, is it used in other contexts?})
% \begin{widetext}
\begin{eqnarray}
\hskip-10mm
\frac{\delta S}{\delta\phi}
=\frac{\partial {\cal L}_{\rm tot.}}{\partial\phi}
-\partial_t
\frac{\partial {\cal L}_{\rm tot.}}{\partial(\partial_t{\phi})}
-\partial_k
\frac{\partial {\cal L}_{\rm tot.}}{\partial(\partial_k\phi)}\nonumber\\&&\hskip-35mm
+\partial_t^2
\frac{\partial {\cal L}_{\rm tot.}}{\partial(\partial_t^2{\phi})}
+2\partial_t\partial_k
\frac{\partial {\cal L}_{\rm tot.}}{\partial(\partial_t\partial_k{\phi})}
+\partial_i\partial_j
\frac{\partial {\cal L}_{\rm tot.}}{\partial(\partial_i\partial_j{\phi})}
=0.\label{EL000}
\end{eqnarray}
% \end{widetext}

Let us illustrate these two approaches in the very simple case of electrostatics where the Lagrangian density reduces to
${\cal L}=\frac 12\varepsilon_0|\vec E|^2-\rho\phi$. Using the first route, the condition $\vec E=-\vec\nabla\phi$ is implemented in the action
through the constraint $S=\int d^3 r\ \![{\cal L}-\vec\lambda\cdot(\vec E+\vec\nabla\phi)]$. Varying the action w.r.t. $\phi$
($\vec E$ being considered as independent of $\phi$)
leads to $\frac{\partial{\cal L}}{\partial\phi}+\vec\nabla\cdot\frac{\partial}{\partial(\vec\nabla\phi)}\vec\lambda\cdot(\vec E+\vec\nabla\phi)=
-\rho+\vec\nabla\cdot\vec\lambda=0$, while variation w.r.t. $\vec E$ produces the equation $\varepsilon_0\vec E-\lambda=\vec 0$. The combination
of the two equations recovers the usual Maxwell equation.
The second approach is more straightforward in this simple case and leads to 
$\frac{\partial{\cal L}}{\partial\phi}-\vec\nabla\cdot\frac{\partial{\cal L}}{\partial(\vec\nabla\phi)}=-\rho-\varepsilon_0\vec\nabla^2\phi=0$, i.e. the Poisson equation.
%\revision{if I understand correctly the higher order derivative here is actually grad phi right?}

The first route is by far simpler in the SO case. 
For the
usual electric and magnetic fields, we obtain the
following equations of motion (Maxwell equations):
% \begin{widetext}
\begin{eqnarray}
\hskip-25mm
\frac{\delta S_{\rm constr.}}{\delta\phi}=
\frac{\delta S_{\rm mat.}}{\delta\phi}+\frac{\delta S_{U(1)}}{\delta\phi}
-\frac{\delta}{\delta\phi}\int dt\ \! d^3r\ \!
\lambda_i^a[\naA_i^a-\frac e{2mc^2}\epsilon_{aji}E_j],
&&\nonumber\\\hskip-15mm
=\frac{\partial{\cal L}_{\rm mat.}}{\partial\phi}-\partial_k
\frac{\partial{\cal L}_{U(1)}}{\partial(\partial_k\phi)}
-\frac e{2mc^2}\partial_k\frac{\partial}{\partial(\partial_k\phi)}(\lambda_i^a
\epsilon_{aji}E_j),
&&\nonumber\\\hskip-15mm
=-\rho_{\rm mat.}+\varepsilon_0\partial_k E_k+{\textstyle\frac e{2mc^2}}\partial_k
\epsilon_{aki}\lambda^a_i,
&&\nonumber\\\hskip-15mm
=0.
\end{eqnarray}
\begin{eqnarray}
\hskip-25mm
\frac{\delta S_{\rm constr.}}{\delta A_i}=
\frac{\delta S_{\rm mat.}}{\delta A_i}+\frac{\delta S_{U(1)}}{\delta A_i}
-\frac{\delta}{\delta A_i}\int dt\ \! d^3r\ \!
\left[\mu^a(\naphi^a+\frac em B^a)
+\lambda_k^a(\naA_k^a-\frac e{2mc^2}\epsilon_{ajk}E_j)\right],
&&\nonumber\\\hskip-15mm
=\frac{\partial{\cal L}_{\rm mat.}}{\partial A_i}-\partial_k
\frac{\partial{\cal L}_{U(1)}}{\partial(\partial_k A_i)}-\partial_t
\frac{\partial{\cal L}_{U(1)}}{\partial(\partial_t A_i)}
+\frac em\partial_k\frac{\partial}{\partial(\partial_k A_i)}(\mu^a B^a)
-\frac e{2mc^2}\partial_t\frac{\partial}{\partial(\partial_t A_i)}(\lambda_k^a
\epsilon_{ajk}E_j),
&&\nonumber\\\hskip-15mm
=J_i-\varepsilon_0 c^2\epsilon_{ijk}\partial_j B_k
+\varepsilon_0\partial_t E_i-{\textstyle\frac em}\epsilon_{ika}\partial_k\mu^a
+{\textstyle\frac e{2mc^2}}\partial_t\epsilon_{ika}
\lambda^a_k,
&&\nonumber\\\hskip-15mm
=0.
\end{eqnarray}
% \end{widetext}
In the first of these expressions, we can define a 
polarization  in terms of the
Lagrange multipliers, 
\begin{equation}
P_k=\frac {e}{2mc^2}\epsilon_{aki}\lambda^a_i.
\end{equation}
Note that the origin of the dielectric polarization has to be found in the
fact that a moving magnetic moment $\vec\mu$ creates a dipolar moment
$\sim\vec v\times\vec\mu$~\cite{KrotkovEtAl}.
This dielectric
polarization is associated to bound charges and
there appears then a 
contribution $-\partial_k P_k$ to the total charge density. This polarization 
also 
contributes the second equation of motion through a term added to the
ordinary current density, $\partial_t P_i$, as well as an additional term
$\epsilon_{ika}\partial_k m^a$, with 
\begin{equation}m^a=-\frac em \mu^a,\end{equation} describing
``Amperian'' currents {(regular charge currents)}.

The electric charge and current densities are
modified by the SO$+$Zeeman terms and the usual Maxwell 
equations become:
\begin{eqnarray}
&&\vec\nabla\cdot{\vecc E}=
\frac1{\varepsilon_0}(\rho-\vec\nabla\cdot{\vecc P}),\\
&&\vec\nabla\times{\vecc B}
-\varepsilon_0\mu_0\partial_t{\vecc E}
=\mu_0[{\vecc J}+\partial_t{\vecc P}+\vec\nabla\times{\vecc m}].
\end{eqnarray}

The variation with respect to the non-Abelian gauge fields leads to 
% \begin{widetext}
\begin{eqnarray}
\frac{\delta S_{\rm constr.}}{\delta\naphi^a}&=&
\frac{\delta S_{\rm mat.}}{\delta\naphi^a}
+\frac{\delta S_{SU(2)}}{\delta\naphi^a}
-\frac{\delta}{\delta\naphi^a}\int dt\ \! d^3r\ \!
\mu^b(\naphi^b+\frac em B^b),\nonumber\\
&=&\frac{\partial{\cal L}_{\rm mat.}}{\partial\naphi^a}
+\frac{\partial{\cal L}_{SU(2)}}{\partial\naphi^a}
-\partial_k
\frac{\partial{\cal L}_{SU(2)}}{\partial(\partial_k\naphi^a)}
-\mu^a,\nonumber\\
&=&-\rhospin_{\rm mat.}^a+{\textstyle \frac\hbar{2c}}\epsilon_{abc}\naA^b_k\naE^c_k
+{\textstyle\frac\hbar c}\partial_k\naE^a_k-\mu^a
\nonumber\\
&=&0.
\end{eqnarray}
\begin{eqnarray}
\frac{\delta S_{\rm constr.}}{\delta\naA^a_i}&=&
\frac{\delta S_{\rm mat.}}{\delta\naA^a_i}
+\frac{\delta S_{SU(2)}}{\delta\naA^a_i}
-\frac{\delta}{\delta\naA^a_i}\int dt\ \! d^3r\ \!
\lambda_k^b(\naA_k^b-\frac e{2mc^2}\epsilon_{bjk}E_j),%\right]
\nonumber\\
&=&\frac{\partial{\cal L}_{\rm mat.}}{\partial\naA^a_i}
+\frac{\partial{\cal L}_{SU(2)}}{\partial\naA^a_i}
-\partial_k
\frac{\partial{\cal L}_{SU(2)}}{\partial(\partial_k\naA^a_i)}
-\partial_t
\frac{\partial{\cal L}_{SU(2)}}{\partial(\partial_t\naA^a_i)}
-\lambda^a_i,\nonumber\\
&=&J_i^a-{\textstyle \frac\hbar{2c}}
\epsilon_{abc}[\naphi^b\naE^c_i+c^2\epsilon_{ijk}
\naA^b_j\naB^c_k]
-\hbar c\epsilon_{ijk}\partial_j\naB^a_k+{\textstyle\frac\hbar c }\partial_t\naE_i^a
-\lambda^a_i,
\nonumber\\
&=&0.
\end{eqnarray}
% \end{widetext}
We recover the radiation spin density
$\rhospin_{\rm rad.}^a=-\frac\hbar{c}\epsilon_{abc}\naA^b_k\naE^c_k$ and the
radiation spin current density 
${\cal J}^a_i=-\frac\hbar{c}\epsilon_{abc}[\naphi^b\naE^c_i+\frac 12
c^2\epsilon_{ijk}
\naA^b_j\naB^c_k]$ in terms of which the
 corresponding Yang-Mills-Maxwell equations become
\begin{eqnarray}
&&\vec\nabla\cdot\vec{\naE}^a=\frac c\hbar(\rhospin^a_{\rm mat.}+\rhospin^a_{\rm rad.}+\mu^a),\label{eqYMM1}\\
&&\vec\nabla\times{\vecc\naB}^a
-\frac 1{c^2}\partial_t{\vecc \naE}^a
=\frac{1}{\hbar c}({\vecc J}^a+\vec{\ \!\cal J}^a-\vec\lambda^a).\label{eqYMM2}
\end{eqnarray}

The non Abelian equations without sources take the form (obtained 
directly from
the 
definitions~(\ref{eqnaE}) and (\ref{eqnaB})) 
\begin{eqnarray}
&&\vec\nabla\times\vec\naE^a+\partial_t\vec\naB^a=
\epsilon_{abc}(\vec\nabla\times (\naphi^b\vec\naA^c)
+{\scriptstyle \frac 12}\partial_t(\vec\naA^b\times\vec\naA^c)),\label{ym3}\nonumber\\
&&\vec\nabla\cdot\vec\naB^a=
{\scriptstyle \frac 12}
\epsilon_{abc}\vec\nabla\cdot(\vec\naA^b\times\vec\naA^c)
.\nonumber 
\end{eqnarray}

%\subsection{Conservation equation}
From the Yang-Mills-Maxwell equations, one can form a continuity equation
describing the conservation of the total angular momentum 
density. Taking the
divergence of Eq.~(\ref{eqYMM2}), and using
Eq.~(\ref{eqYMM1}), one gets
$$\partial_t(\rhospin^a_{\rm mat.}+\rhospin^a_{\rm rad.})+\vec\nabla\cdot(\vec
J^a+\vec{\ \!\cal J}^a)=-\partial_t\mu^a+\vec\nabla\cdot\vec\lambda^a.$$
The Lagrange multipliers formally appear in the continuity equation, but 
physically, the l.h.s. contains all  sources of angular 
momentum incorporated in the problem~\cite{BercheEtAl12}, i.e. 
from free electrons in the conduction band encoded in the 
spinor $\psi$ (the matter spin density),
and from angular momentum transfered to the 
lattice via SO interaction (the radiation angular momentum density).  
So the conservation of angular momentum in the system reads as expected
\begin{equation}
\partial_t(\rhospin^a_{\rm mat.}+\rhospin^a_{\rm rad.})+\vec\nabla\cdot(\vec
J^a+\vec{\ \!\cal J}^a)=0.\label{eqnConservation}\end{equation}
We note that in the case of a static and homogeneous problem, the
Lagrange multipliers just take the simple form
$\mu^a=-(\rhospin^a_{\rm mat.}+\rhospin^a_{\rm rad.})$ and 
$\vec\lambda^a=\vec J^a+\vec{\ \!\cal J}^a$.

\section{Einstein - de Haas experiment revisited}\label{secVI}
The Einstein-de Haas experiment is one of the famous experiments of the
beginning of the XXth century based on gyromagnetic 
phenomena~\cite{Richardson1908,Barnett1935,Frenkel79}. In this
experiment, a cylinder made of a non magnetized ferromagnetic material
is suspended to a torsion wire and an external magnetic field is applied 
along the cylinder's axis. The cylinder acquires a magnetization, {because spins orient to
the field lowering the energy of the system} and at the
same time, the cylinder {starts a rotating motion}.
This effect became an efficient 
experimental method for the measurement of the gyromagnetic ratio of various 
materials and, for example, proved that magnetism of iron is essentially due
to the spin degrees of freedom.

This experiment is well known, and well understood from very fundamental 
principles, since it relies on the conservation of angular momentum.
When the material gets magnetized, the individual magnetic moments 
of the electrons point in a common direction, hence a total angular momentum
appears which has to be compensated by an opposite angular momentum carried
by the whole sample in order to conserve the initially vanishing value of the
total angular momentum of the sample. 
Although this mechanism is very clear and unquestionable, there is no simple
microscopic explanation of how this conservation of 
angular momentum is at play at the atomic level. See 
however\cite{JaafarEtAl09} 
for a beautiful description of the dynamics of the effect in terms 
of elastic twist on the lattice induced by the rotation of localized spins
(spin-rotation coupling) and the coupling to the anisotropy 
field. {The latter anisotropy is in fact a consequence
of the existence, at a deeper level, of spin-orbit interactions.
The purpose of this section is to provide such a microscopic description}. Our 
approach is based on the gauge field description of spin-orbit
interactions. The interest of the gauge field theory 
here is to provide a natural way for the introduction of conserved 
currents carrying angular momentum.

An external magnetic field $\vecc B=B\vecc e_z$ is applied 
to the cylinder
of ferromagnetic material initially non magnetized, suspended along its
$z-$axis to a torsion wire
(hence the appearence of a non-Abelian scalar field $\naphi^3$ in the 
Maxwell-Yang-Mills approach). The material gets magnetized. The spins acquire
a common orientation and the corresponding magnetization is called the matter
spin polarization. 
The sample is subject to a torsion around the vertical axis, and the 
radiation counterpart is the angular momentum acquired by the lattice.
We can understand the physical situation as the appearence of a matter 
spin polarization $\rhospin_{\rm mat.}^3$ since the individual localized
spins get polarized
along the direction of the external magnetic field. As a consequence,
$\vec\nabla\cdot\vec\naE^3\not =0$ and  the 
Yang-Mills electric field aquires a non vanishing component $\vec\naE^3$.
The localized spins interact with the lattice ions through an SO interaction
(responsible for example for crystalline anisotropy), hence some components
of $\vec\naA^a$ do not vanish, creating, together with $\naphi^3$ a radiation
contribution to the spin polarization 
$\rhospin_{\rm rad.}^3\sim\varepsilon_{3bc}\vec\naA^b\cdot\vec\naE^c$.

Let us analyse the equations in order to see whether an equation of rotation 
of the lattice follows. Here, the lattice enters the problem
via the SO interaction, 
i.e. via the electric field (the ``true'' internal electric field) which is ``rigidly'' 
associated to the lattice ions. 
Since
$\rhospin_{\rm rad.}^3=-(\hbar/c)\varepsilon_{3bc}\naA^b_k\naE^c_k$,
and $\naE^c_k=-\partial_t\naA^c_k+\epsilon_{cde}\naphi^d\naA^e_k$ (we assume
here that $\naphi^d=\naphi^3$, determined by the applied
magnetic field, is uniform)
 we obtain
\begin{equation}\rhospin_{\rm rad.}^3={\textstyle\frac{e^2\hbar}{4m^2c^5}}
(\vec E\times\partial_t\vec E+{\textstyle\frac em}(|\vec E|^2+(E_3)^2)\vec B)_3.
\end{equation}
As we had noticed, the appearence of the radiation contribution to
the polarization is connected to the SO interaction (hence the electric field).
This equation self-consistently shows
that the true electric field produced by the ions
located at the lattice sites is subject to a motion of rotation around the
axis $\vecc e_3$ along which the external magnetic field was applied.
This
rotation of the electric field can only be caused by an equivalent rotation of 
the whole lattice, thus explaining the observed Einstein-de Haas effect. 
If we now take into account the magnetization acquired by the cylinder
and
 the exchange interaction between the conducting electrons and the localized moments ($J_{sd}$ coupling constant), 
$\vec B$ has to be changed into $\vec B_{\rm eff}=\vec B+J_{sd}\mu_0\vec M$.

\section{Conclusion}\label{secIX}
In this paper, we have shown that spin-orbit interaction  
is a suitable arena to introduce non-Abelian gauge field theory
(namely here SU(2) gauge theory) and that many consequences, relevant in 
a condensed matter physics context can be drawn from
the classical form of the theory.
More specifically, we have discussed the conserved current associated to this 
theory, the equations of motion of the gauge field, 
and the role of the continuity 
equation in the discussion of the Einstein - de Haas experiment. 
%and we have shown how the Landau-Lifshitz relaxation torque and the Gilbert damping follow from a general conservation equation of the total angular momentum carried by both the spin current and the lattice ions via the SO interaction. 
The approach follows from a very general
Yang-Mills-Maxwell theory which we believe is likely suitable for
micromagnetic simulations of spin current and spin configurations in
various materials such as conducting ferromagnets or 2DEGs.

\acknowledgements This work was supported by CNRS-Fonacit grant PI-2008000272, and by the support of Nancy-Universit\'e through an Invited Professor position (EM). EM also acknowledges support from the POLAR Foundation.

\section*{References}
\def\paper#1#2#3#4#5{{#1,}{\ {\it #2}}{\ {\bf #3},}{\ #4}{\ (#5).}}
\def\papertitle#1#2#3#4#5#6{{#1,}{\ ``#6,''}{\ {\it #2}}{\ {\bf #3},}{\ #4}{\ (#5).}}

\end{document}